# Exciton-exciton Interaction in Monolayer MoSe$_2$ from Mutual Screening of Coulomb Binding

Ke Xiao[1,6], Tengfei Yan[2], Chengxin Xiao[1,7], Feng-ren Fan[1,7], Ruihuan Duan[3], Zheng Liu[3], Kenji Watanabe[4], Takashi Taniguchi[5], Wang Yao[1,7] *, Xiaodong Cui[1] *

1 Department of Physics, University of Hong Kong, Hong Kong SAR

2 School of Microelectronics, Shanghai University, Shanghai, China

3 School of Materials Science and Engineering, Nanyang Technological University, Singapore 639798, Singapore

4 Research Center for Functional Materials, National Institute for Materials Science, 1-1 Namiki, Tsukuba 305-0044, Japan

5 International Center for Materials Nanoarchitectonics, National Institute for Materials Science, 1-1 Namiki, Tsukuba 305-0044, Japan

6 NISE Department, Max Planck Institute of Microstructure Physics, Halle, Germany

7 HKU-UCAS Joint Institute of Theoretical and Computational Physics, Hong Kong SAR

**Abstract**

The potential for low-threshold optical nonlinearity has received significant attention in the fields of photonics and conceptual optical neuron networks. Excitons in two-dimensional (2D) semiconductors are particularly promising in this regard as reduced screening and dimensional confinement foster their pronounced many-body interactions towards nonlinearity. However, experimental determination of the interactions remains ambiguous, as optical pumping in general creates a mixture of excitons and unbound carriers, where the impacts of band gap renormalization and carrier screening on exciton energy counteract each other. Here by comparing the influences on exciton ground and excited states energies in the photoluminescence spectroscopy of monolayer MoSe$_2$, we are able to identify separately the screening of Coulomb binding by the neutral excitons and by charge carriers. The energy difference between exciton ground state (*A-1s*) and excited state (*A-2s*) red-shifts by 5.5 *meV* when the neutral exciton density increases from 0 to $4 \times 10^{11} cm^{-2}$, in contrast to the blue shifts with the increase of either electron or hole density. This energy difference change is attributed to the mutual screening of Coulomb binding of neutral excitons, from which we extract an exciton polarizability of $\alpha_{2D}^{exciton} = 2.55 \times 10^{-17} eV \left(\frac{m}{V}\right)^2$. Our finding uncovers a new mechanism that dominates the repulsive part of many-body interaction between neutral excitons.

**Introduction**

Excitons are neutral quasiparticles that can be free of permanent electrical dipole and multipoles as well. Coulomb interactions of its electron and hole constituents, on the other hand, give rise to many-body interactions of these composite bosons which can take various forms.[1-3] These complicated interactions dynamically modify the exciton resonance energy and potentially lead to optical nonlinearity. Excitons also interact with electrons or holes in doped systems, which can lead to renormalization of band gap and screening of Coulomb binding,[4,5] and formation of Fermi polaron, [6-10] etc.

Monolayer transition metal dichalcogenides (TMDs) have provided a platform to explore exciton phenomena in the two-dimensional (2D) limit. Owing to quantum confinement and the reduced dielectric screening in 2D, excitons in monolayer TMDs exhibit giant binding energy with energetically well separated Rydberg states, [11-16] promising the exploration of excitonic many-body phenomena and optoelectronic applications in ambient conditions.

One standing question regarding 2D excitons is how the exciton resonant energy is affected by the various factors arising from the enhanced Coulomb interaction, which normally coexist in experimental systems.[4] The static screening effect arising from the environmental susceptibility has been recognized as a control to tune the exciton properties with various approaches of dielectric engineering.[17-24] Compared to dielectric environments, doping the monolayer with a bath of electrons/holes,[22] or even excitons, can have a greater impact on the Coulomb interaction and consequently the exciton binding energy. Intuitively, charge carriers and charged excitons (trions) can screen Coulomb interaction, reducing the exciton binding energy, and tend to blueshift exciton resonance. In the meantime, the quasiparticle band gap gets renormalized which tends to redshift exciton resonance.[25-31] Distinguishing the competing factors in a quantitative manner therefore remains an experimental challenge. Furthermore, even neutral excitons may play a role in Coulomb screening like polarizable atoms or molecules, and affects binding energies of each other. Such effect, however, remains unexplored.

In this letter we report the experimental determination of the mutual screening of Coulomb binding of neutral excitons in high quality monolayer $MoSe_2$ with photoluminescence spectroscopy. We utilize the energy difference $\Delta E = E_{1s} - E_{2s}$ between the 1s exciton ground state and the 2s excited Rydberg state to unambiguously monitor the change of exciton binding energy, where the contribution of band gap normalization can be completely excluded. From the narrow spectra resonances, we observe a clear red shift of $\Delta E$ by 5.5 $meV$ when the neutral exciton density increases from 0 to $4 \times 10^{11} cm^{-2}$, while the trion density remains negligibly low. In contrast, increasing electron or hole density leads to a blue shift in $\Delta E$. These opposite trends unambiguously suggest that the electron-hole binding in an exciton can be appreciably screened by the surrounding excitons, through inducing electrical polarization in these neutral quasiparticles. The exciton polarizability extracted from our PL measurements, $\sim 2.55 \times 10^{17} eV \left(\frac{m}{V}\right)^2$, is in excellent agreement with the nonlinear Stark shift measured in applied electric fields.[32-34] This realizes a repulsive many-body interaction of neutral excitons which has a dominating strength as compared to the exciton interactions from Coulomb exchange [1,35], whereas its sensitive dependence on the exciton Rydberg orbitals further distinguishes it from the effective attractive part from bandgap renormalization.

## Results

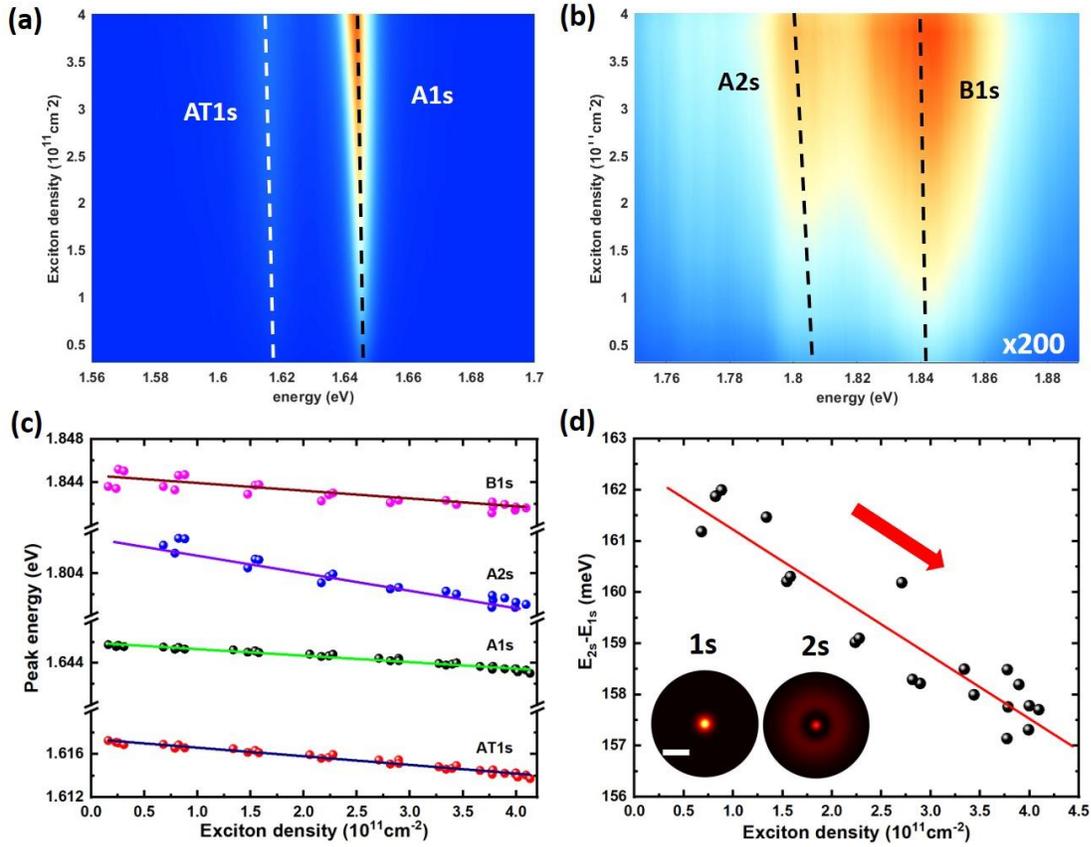

**Figure 1.** The intensity-dependent PL spectral map of excitons (a) ground state of band-edge exciton (*A-1s*) and the corresponding charge bound exciton (*AT-1s*), (b) the first excited state of band edge exciton (*A-2s*) state and the ground-state spin-off exciton (*B-1s*). The PL intensities of *A-2s* state and *B-1s* are magnified by 200X for better comparison. The trends of exciton energy indicated by the dashed lines show redshift under the increased excitation intensity. (c) The peak energy shifts of *A-1s*, *AT-1s*, *B-1s* and *A-2s* states *v.s.* the exciton density (c.f. Table S1 in SI for the quantitative estimation). (d) The energy difference $\Delta E$ between *A-1s* and *A-2s* states shrinks at the elevated exciton density. The inset shows the exciton wavefunction of *A-1s* and *A-2s* states in real space. The scare bar represents 4nm.

Figure 1 summarizes the exciton-density dependent photoluminescence spectra under the excitation of 2.331eV. The prominent PL peaks around 1.646eV and 1.618eV are assigned to the band edge exciton *A-1s* and its trion *AT-1s* (charge bounded exciton). The two weak PL peaks (magnified X200 with respect to that of *A-1s*) at ~1.808eV and ~1.844eV are attributed to the first excited state of *A* exciton, labelled as *A-2s* and *B* exciton, labelled as *B-1s*, respectively, according to ref [8,9]. Here *A* exciton and *B* exciton originate from the spin splitting of the conduction band and valence band at *K(K')* valley where the direct band gap is located.[36] The PL intensities of *A-2s* and *B-1s* are observed two orders of magnitude weaker than that of *A-1s* since both are either the excited state or not the band edge excitons. The ground state exciton and its trion (*A-1s*, *AT-1s*) undergo an apparent quantum yield reduction with increasing excitation power (Fig.S2), which may be attributed to Auger recombination. [37] The excitation power dependent PL measurement is then utilized to estimate the exciton density. (Supplementary Note 2). More interestingly, all the excitons and trion (*A-1s, AT-1s, B-1s, A-2s*) undergo an obvious redshift

(Fig.1(a-b)) at the elevated exciton density. Specifically, the exciton energy shift determined from the Lorentzian fitting of the PL spectra are summarized in Fig.1(c). The energy shifts of *A-1s, AT-1s* and *B-1s* show a similar dependence on the exciton density. Contrarily, the peak energy of *A-2s* redshifts obviously at a steeper slope than that of ground state (Table.S1).

Usually, the bandgap renormalization induced by photo doping leads to a redshift of the quasiparticle bandgap, whereas the screening effect induced by photo doping results in the decrease of exciton binding energy and consequently leads to an energy blueshift. Therefore, from our experimental results in monolayer MoSe$_2$ (Fig.1c), the bandgap renormalization effect appears to be dominant as all exciton states observed (*A-1s, AT-1s, B-1s, A-2s*) experience a redshift at the elevated excitation intensity.

It is worth mentioning that the bandgap renormalization has the same influence on *A-1s* and *A-2s* excitons since the electrons and holes of *A-1s* and *A-2s* excitons come from the same band edges. Therefore, the energy difference between *A-1s* and *A-2s* states are solely dependent on the exciton binding energy. Fig.1(c) shows that the energy difference between *A-1s* and *A-2s* states gradually shrinks with the elevated excitation intensity. The different energy shift slope (Fig.1(c)) and the shrinking energy difference may imply that the ground (1s) and the first excited states (2s) experience screening effect to a different extent due to their different Bohr radius ($r_B^{1s} \sim 1nm$ v.s. $r_B^{2s} \sim 3nm$).

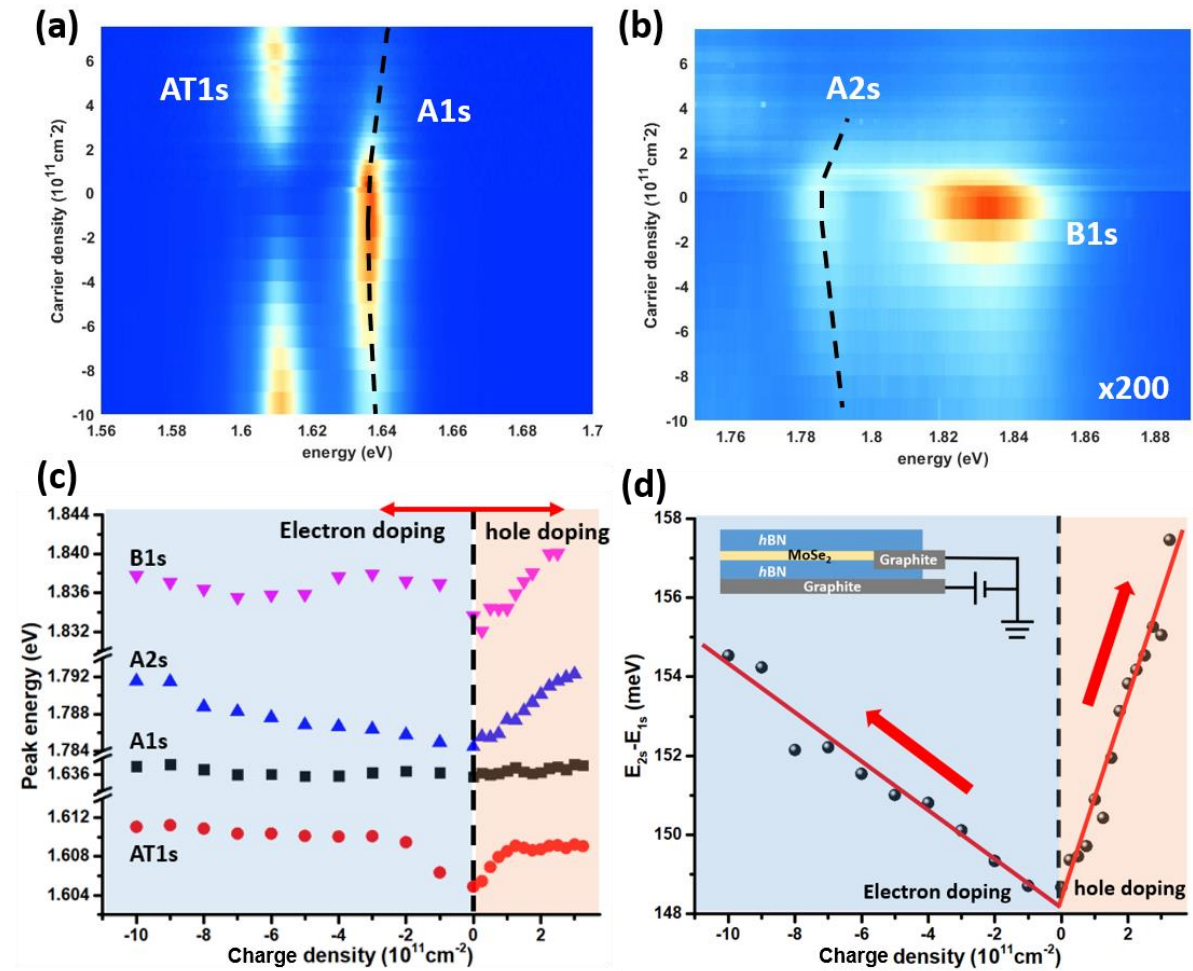

**Figure 2.** Charge-density dependent photoluminescence spectra of (a) *A-1s* and *AT-1s* and (b) *A-2s* and *B-1s* excitons at 10K. The PL intensities of *A-2s* state and *B-1s* are magnified by 200X for better comparison. (c) The peak energy of *A-1s*, *AT-1s*, *B-1s*, *A-2s* states as a function of the electron and hole density. (d) The energy difference ΔE between *A-1s* and *A-2s* excitons displays clearly blue-shift at the elevated electron/hole density, contrasting to red-shift at the increased exciton density. The inset figure shows the schematic device structure in the carrier-density dependent PL measurement. The charge density is tuned by an electrostatic gating via *h-BN* cap layer (~20nm thick). Paste figure above the legend.

In principle, the photo excitation can inject unbound photo carriers in addition to the neutral excitons. To distinguish their effects on the different energy shifts of *A-1s* and *A-2s*, we conduct a series of PL experiments with electrostatic doping. Figure 2 summarizes our carrier-density dependent PL results. At the elevated carrier density, both *A-1s* and *A-2s* undergo a blue shift (Fig.2(a-b)), which is consistent with the previous reports. [23] We attribute the blue shift to the exciton-polaron effect based on many-body charge-exciton interaction.[6,8,38] On the other hand, $\Delta_{2s-1s}$, the energy difference between *A-1s* and *A-2s* states increases at elevated carrier densities as Fig.2(c) shows. This contrasting shift of $\Delta_{2s-1s}$ as functions of carrier density (positive slope) *vs.* the experimentally observed negative slope with exciton intensity ambiguously suggests that the latter is not from the unbound photo carriers. This is further corroborated by the weak trion emission (Fig.1(a)) in comparison with that of *A-1s* throughout the entire range of excitation intensity, which has implied a very low density of the unbound photocarriers.

The above analysis points to interaction between neutral excitons as the cause of $\Delta_{2s-1s}$ decreasing with the excitation intensity. As we detail below, while these excitons do not carry charge and dipole at rest, the Coulomb interaction between the electron and hole constituents can polarize adjacent excitons, which in turn screens the Coulomb and leads to reduction in the binding energy. The opposite trends of the charge- and exciton-density dependence in the *1s -2s* energy difference suggests that our measured redshift under the increased exciton density provides a lower bound for this mutual screening effect among the neutral excitons.

To quantitatively examine the neutral exciton mutual screening effect, we numerically solve the Schrodinger equation of exciton binding

$$\left(-\frac{\hbar^2}{2\mu}\nabla^2 + V(r)\right)\varphi_l = E_l\varphi_l$$

with the well-established Rytova-Keldysh form [39,40] ($V(r) = -\frac{e^2}{8\varepsilon_0 r_0}\left[H_0\left(\frac{\kappa r}{r_0}\right) - Y_0\left(\frac{\kappa r}{r_0}\right)\right]$) which describes the screened Coulomb interaction in two-dimensional geometry with parameters of dielectric constant ($\kappa$) and screening length ($r_0$). The screening length $r_0$ accounts both the in-plane electric polarizability of pristine monolayer TMDs and that from the polarizability of the neutral exciton bath (Supplementary Note 6). Hence, we fix the effective reduced mass ($\mu = 0.27 m_e$) and dielectric constant ($\kappa = 4.5$) according to ref [41] while setting the screening length as the sole variable to reflect the change of dielectric environment with exciton density.

The solution indicates that the binding energy of the ground state and the excited-states excitons are influenced by the screening length to different extents, as summarized in Fig.3(a). It is noted that the energy shift of the higher excited state is less sensitive to the variation of the screening length, namely $\frac{\Delta E_{2s}}{\Delta r_0} < \frac{\Delta E_{1s}}{\Delta r_0}$. This could be understood in the way that the wavefunction of *A-1s* state has much smaller radius than that of *A-2s* state (c.f. inset in Fig.1(d)), and consequently is influenced more by the change of screening length in the Keldysh potential (Supplementary Note 8). By fitting our exciton-density dependent PL results (Fig.1(d)) with the model, we can extract the screening length as a function of the exciton density as shown in Fig.3(b). (Supplementary Note 7) It implies that a change of the exciton density by $\sim 4.5 \times 10^{11} cm^{-2}$ effectively tunes the screening length by $\Delta r_0 \sim 0.25$nm at $r_0 \sim 4.2 nm$.

The screening length is linearly proportional to the in-plane electric polarizability of pristine monolayer TMDs and that of the injected neutral exciton bath: $r_0 = \frac{1}{2\kappa\varepsilon_0}(\alpha_{2D}^{MoSe_2} + n\alpha_{2D}^{exciton})$, from which we could extract the pristine 2D electric polarizability of monolayer MoSe$_2$: $\alpha_{2D}^{MoSe_2} = 3.28 \times 10^{-19} \left(\frac{C}{V}\right)$, and the 2D exciton polarizability $\alpha_{2D}^{exciton} = 2.55 \times 10^{-17} eV \left(\frac{m}{V}\right)^2$. This in-plane exciton polarizability agrees well with the previous calculations [32] and experimentally measured values from the nonlinear Stark effect in applied electric field [33,34,42], which further confirms that the exciton-density dependent screening effect originates from excitons polarizability.

Our results show that, in analogy to the carrier screening effect (Fig.3(c)-middle) which modifies the effective Coulomb potential, neutral exciton can also induce a screening effect with a relatively modest but clearly visible contribution (Fig.3(c)-right). Fig.3(b) plots the calculated binding energies of *A-1s* and *A-2s* excitons as functions of the exciton density. Together with the measured *A-1s* and *A-2s* exciton resonances, we can also extract the bandgap renormalization as a function of the exciton density, which is well consistent with the photoinduced bandgap renormalization measurements [25,26] and calculations [43,44].

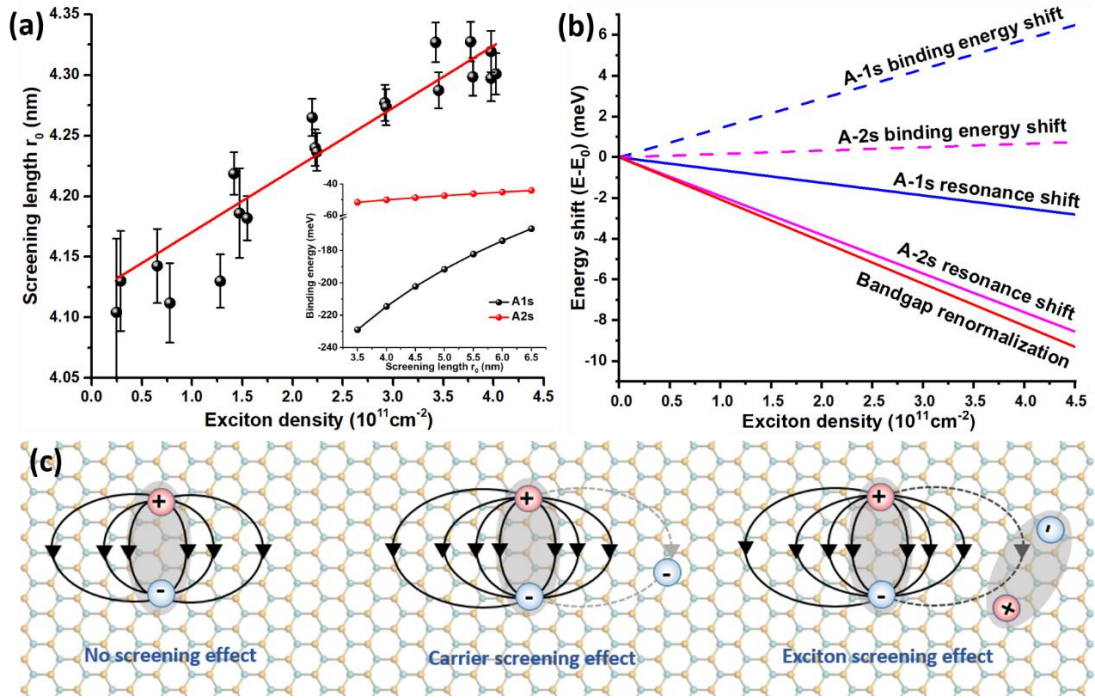

**Figure 3.** (a) The screening length fitted from the experimental data as a function of the exciton density. Inset: the calculated binding energy variation of *A-1s, A-2s* exciton states as functions of the screening length ($r_0$). (b) The extracted binding energy change of *A-1s*, *A-2s* and bandgap renormalization as functions of exciton density. (c) Schematics of charge screening and exciton polarization. The dashed line

represents the charge or dipole screening effect. The more accurate description could be found at supplementary information.

**Discussion**

The mutual screening of neutral excitons effectively realizes a repulsive interaction between these composite bosons. From the linear dependence on exciton density (c.f. Fig. 1 and 3), we can write the exciton Hamiltonian as $H_X = \sum_{l,k} E_{l,k}(n) \hat{X}_{l,k}^\dagger \hat{X}_{l,k}$, $E_{l,k}(n) = E_{l,k}(0) + \eta_l n$, where $E_{l,k}(0)$ is the exciton dispersion of Rydberg state $l$, $n = \langle \sum_{l,k} \hat{X}_{l,k}^\dagger \hat{X}_{l,k} \rangle$ is the exciton density (predominantly the 1s exciton). $\eta_l$ are the slopes of the dashed lines in Fig. 3(b), which correspond to the interaction strength of exciton in Rydberg state $l$ with the bath of 1s excitons. In a mean-field description, this repulsive interaction can be written as $H_{int} = \sum_{l,k} \eta_{l,1s} \hat{X}_{l,k}^\dagger \hat{X}_{l,k} \langle \sum_{k'} \hat{X}_{1s,k'}^\dagger \hat{X}_{1s,k'} \rangle$. Our measurements determined the effective interaction strength of 1s excitons $\eta_{1s,1s} = 1.44 \times 10^{-11} meV \cdot cm^2$, and between 1s and 2s excitons $\eta_{2s,1s} = 1.66 \times 10^{-12} meV \cdot cm^2$. In the literature, Coulomb exchange is generally considered as the dominant cause for the repulsive part of interaction between neutral excitons [1,35]. We find that $\eta_{1s,1s}$ from this mutual screening is one order of magnitude larger as compared to that of the exchange exciton-exciton interaction (c.f. Supplementary Note 9 and Fig.S8). So in monolayer TMDs, this mutual screening mechanism becomes the dominant contribution to the repulsive part.

In summary, we investigate the energy difference between excitonic Rydberg energies in monolayer MoSe$_2$ as a function of charge and exciton densities by photoluminescence spectroscopy. This energy difference reflects the exciton binding energy and is immune to the band gap renormalization. The contrasting dependence of the energy difference between *A-1s* and *A-2s* states on the exciton density (negative slope) *v.s.* charge density (positive slope) supports that the screening effect originates from the polarizability of the neutral excitons. With the model of Rytova-Keldysh potential, the exciton polarizability is extracted to be $2.55 \times 10^{-17} eV \left(\frac{m}{V}\right)^2$ up to the exciton density of $4.5 \times 10^{11} cm^{-2}$, well consistent with previous experimental results from nonlinear Stark effect in in-plane electric field. This mutually screening effect between the excitons effectively manifests as a new many-body interaction between the tightly bound excitons in the two-dimensional geometry. This finding calls for microscopic formulation of the exciton-exciton interaction beyond the existing framework that treats the Rydberg orbitals as rigid ones.

**Materials and Methods**

Crystal growth:

Bulk MoSe$_2$ crystals are grown using the chemical vapor transport (CVT) method. Silica tubes are loaded with Mo powder (99.9%), slightly excessive Se ingot (99.999%), and a small amount of iodine as transport agents. The tubes are then evacuated and sealed. Next, the silica tubes are placed in the reaction zone at 950 ℃ and the growth zone at 900 ℃. After a duration of fifteen days, large-sized bulk MoSe$_2$ crystals are obtained in the cold zone.

Sample preparation:

The monolayer MoSe$_2$ and few-layer h-BN are mechanically exfoliated onto a Si substrate with a 285 nm SiO$_2$ film. Subsequently, the monolayer MoSe$_2$ is encapsulated by h-BN using the dry transfer method. (43) For the device utilized in carrier density-dependent PL measurement, the dry-transfer method is employed to stack different 2D materials in the sequence of h-BN/Graphite/MoSe$_2$/h-BN/Graphite. Additionally, few-layer graphite is employed to establish better contact between the sample and the Au electrode, as illustrated in Figure S1.

Exciton density dependent PL measurement:

PL spectroscopy was performed based on a home-made confocal microscopy system, using solid state 532nm continuous laser (*Excelsior 532, Spectra-Physics*). Signals were collected in the reflection configuration via a notch filter and dispersed by spectrograph (*Shamrock 193i*) prior to detection with an built-in EMCCD (*Andor*). Half-wave plate and polarized beam splitter cube were used to change the excitation power automatically with a program to avoid the artificial factors.

**Acknowledgments**

The work was supported by the National Key R&D Program of China (2020YFA0309600), Guangdong-Hong Kong Joint Laboratory of Quantum Matter and the University Grants Committees/Research Grants Council of Hong Kong SAR (AoE/P-701/20, 17300520). K.W. and T.T. acknowledge support from the Elemental Strategy Initiative conducted by the MEXT, Japan (Grant Number JPMXP0112101001) and JSPS KAKENHI (Grant Numbers 19H05790, 20H00354 and 21H05233). R.D and Z.L. acknowledge support from the Singapore Ministry of Education Tier 3 Programme "Geometrical Quantum Materials" AcRF Tier 3 (MOE2018-T3-1-002), AcRF Tier 2 (MOE2019-T2-2-105). The authors thank Mr. Mingyang Liu, Dr. Bairen Zhu and Mr. Huiyuan Zheng for fruitful discussion.